\documentclass[conference]{IEEEtran}
\usepackage{cite}
\usepackage{amsmath,amssymb,amsfonts, mathtools}
\usepackage{algorithmic}
\usepackage{graphicx}
\usepackage{textcomp}
\usepackage{xcolor}
\usepackage[normalem]{ulem}
\usepackage{booktabs, makecell, tabularx}
\usepackage[font=scriptsize,labelfont=bf]{caption}
\newcolumntype{Y}{>{\centering\arraybackslash}X}

\graphicspath{{images/}}

\useunder{\uline}{\ul}{}

\def\BibTeX{{\rm B\kern-.05em{\sc i\kern-.025em b}\kern-.08em
    T\kern-.1667em\lower.7ex\hbox{E}\kern-.125emX}}

\begin{document}

\title{Chest X-Rays Image Classification from $\beta$-Variational Autoencoders Latent Features \\
}

\author{\IEEEauthorblockN{Leonardo Crespi}
\IEEEauthorblockA{\textit{\textbf{Dipartimento di Elettronica,}} \\
\textit{\textbf{Informazione e Bioingegneria}} \\
\textit{Politecnico di Milano}\\
Milan, Italy; \\
\textit{\textbf{Centre for Health Data}}\\
\textit{Human Technopole}\\
Milan, Italy \\
leonardo.crespi@polimi.it}
\and
\IEEEauthorblockN{Daniele Loiacono}
\IEEEauthorblockA{\textit{\textbf{Dipartimento di Elettronica,}} \\
\textit{\textbf{Informazione e Bioingegneria}} \\
\textit{Politecnico di Milano}\\
Milan, Italy \\
daniele.loiacono@polimi.it}
\and
\IEEEauthorblockN{Arturo Chiti}
\IEEEauthorblockA{\textit{\textbf{Dipartimento di Medicina Nucleare}} \\
\textit{Humanitas Research Hospital}\\
Milan, Italy\\
arturo.chiti@hunimed.eu}
}

\maketitle

\begin{abstract}
Chest X-Ray (CXR) is one of the most common diagnostic techniques used in everyday clinical practice all around the world. We hereby present a work which intends to investigate and analyse the use of Deep Learning (DL) techniques to extract information from such images and allow to classify them, trying to keep our methodology as general as possible and possibly also usable in a real world scenario without much effort, in the future. To move in this direction, we trained several $\beta$-Variational Autoencoder ($\beta$-VAE) models on the CheXpert dataset, one of the largest publicly available collection of labeled CXR images; from these models, latent features have been extracted and used to train other Machine Learning models, able to classify the original images from the features extracted by the $\beta$-VAE. Lastly, tree-based models have been combined together in ensemblings to improve the results without the necessity of further training or models engineering. Expecting some drop in pure performance with the respect to state of the art classification specific models, we obtained encouraging results, which show the viability of our approach and the usability of the high level features extracted by the autoencoders for classification tasks. 
\end{abstract}


\section{Introduction}
Fuelled by the developments of Convolutional Neural Networks (CNNs)\cite{lecun}, deep learning proved to be a successful approach to tackle challenging problems involving image classification and understanding~\cite{vgg16}.
Thus, the application of these techniques to medical imaging is attracting an increasing amount of interest in the last few years~\cite{greenspan2016guest,ravi2017deep} with the promise of reducing the time and the cost of medical practices.
Unfortunately, to successfully apply deep learning to medical imaging, we still need to deal with many challenges, such has the limited amount of public dataset, the need of expensive computational resources, and the heterogeneity in data acquisition processes.
On the other hand, a major benefit of deep learning over other machine learning approaches is its capability of learning effective representations directly from raw data, instead of relying on feature engineering processes specific for each problem.
Accordingly, in this work we investigate the application of deep learning approaches to extract useful features from medical images that can be later exploited to train classification, segmentation, localisation, and prediction models using eventually more simple and less computationally expensive techniques.
Following the same approach of~\cite{giacomello2021image}, we focused on the problem of chest X-ray (CXR) interpretation, that represents perhaps the most frequent imaging examination performed for screening, diagnostic purposes, and management of many life threatening diseases.
For this reason, in the last few years, several large dataset have been made available for this problem, such as the \emph{CheXpert} dataset\cite{irvin2019chexpert}, which includes more than 200k CXR labeled images and has been used for a scientific competition on automated CXR   interpretation \cite{chexpert_competition}.

In this paper, inspired by our previous work~\cite{giacomello2021image}, we trained several $\beta$-Variational Autoencoder ($\beta$VAE) models on the CheXpert dataset to compute compact image embeddings,  i.e., to extract a set of high-level features from the images.
Then, these image embeddings have been used to train several classifiers (Random Forest, Gradient Boosting, Extremely Randomized Trees and K-Nearest Neighbours) on the problem of finding the five major thoracic diseases (Enlarged Cardiomegaly, Edema, Consolidation, Atelectasis, and Pleural Effusion) in the CXR dataset.
Finally, similarly to what found in~\cite{pham2019interpreting} and in~\cite{giacomello2021image}, we showed that ensemble strategies can be employed to effectively combine several models -- trained with image embeddings computed using different autoencoders -- in order to improve the classification performances.
Our results are promising and show that these models can achieve reasonable performances with respect to the ones achieved by CNNs trained specifically to classify the same images, resulting in a decrease of the AUROC that ranges from around 5\% to 9\%.
These results show that the image embedding models, although not specifically trained to extract features useful to classify the CheXpert images, are nonetheless able to extract relevant and general features of the images that allow to identify the different diseases.

The paper is organised as follows.
In Section~\ref{sec:related} we discuss the most relevant related works.
In Section~\ref{sec:methodology} we provide an overview of the approach proposed in this paper.
Then, Section~\ref{sec:design} we describe the experimental design employed and in Section~\ref{sec:results} we illustrate the results obtained.
Finally in Section~\ref{sec:conclusions} we draw some conclusions and we discuss about possible further developments. 
 \section{Related Works}
\label{sec:related}
Along with the availability of large dataset and with the rise of several related challenges~\cite{johnson2019mimiccxrjpg,irvin2019chexpert,Wang_2017}, several authors successfully applied deep learning and convolutional neural network to chest CXR image understanding.
In~\cite{rajpurkar2017chexnet}, Rajpurkar et al. introduced \emph{CheXNet}, a DenseNet-121~\cite{densenet} network trained on the ChestX-ray14 dataset~\cite{Wang_2017}, achieving state-of-the-art performance on the classification of the 14 thoracic diseases while also outperforming expert radiologists on the detection of pneumonia.
Later, Rajpurkar et al.~\cite{rajpurkar2018deep} extended their previous work introducing \emph{CheXNeXt} improving \emph{CheXNet} and achieving a performance similar to expert radiologist on 10 thoracic diseases.
Other notable works, that exploited the ChestX-ray14 dataset, include the work of Kumar et al.~\cite{kumar2018boosted}, who trained cascaded CNNs to diagnose all the 14 thoracic diseases and the work of Lu et al.~\cite{Lu2020Multi}, who exploited evolutionary algorithm to search for the best CNN architecture to classify the images.
An additional notable work is the one of Ye et al.~\cite{ye2020weakly} that proposed \emph{Probabilistic-CAM} (PCAM), an extended version of Class Activation Mapping (CAM)~\cite{cam_paper}, that allows to localize thoracic diseases on the ChestX-ray14 dataset in a semi-supervised fashion;
moreover, their localization model has been also successfully applied to image classification, outperforming some of the approaches previously introduced in the literature.
Recently, two very large datasets have been released: CheXpert~\cite{irvin2019chexpert} and MIMIC-CXR~\cite{johnson2019mimiccxrjpg}, consisting of respectively 224000 and 350000 images.
Exploiting these datasets, Irvin et al.~\cite{irvin2019chexpert} designed a classifier based on a 121-layer DenseNet trained with different strategies to deal with uncertainty that is present in the labels of CheXpert dataset.
This model achieved a performance level comparable to that of an expert radiologist on the classification of the 5 most representative thoracic diseases.
Rubin et al.~\cite{rubin2018large} introduced \emph{DualNet}, which consists of two CNNs jointly trained on frontal and lateral chest radiography, included in the very large MIMIC-CXR dataset~\cite{johnson2019mimiccxrjpg}.
Their results suggest that classifiers trained on two types of images (i.e., either frontal or lateral), like \emph{DualNet}, are able to outperforms state-of-the-art classifiers trained separately on a single type of image.
Instead, Pham et al.\cite{pham2019interpreting} trained several CNNs on the CheXpert dataset, showing the benefits of exploiting the conditional dependencies among the labels, resulting from their hierarchy, in the training as well as the possibility of exploiting classifiers ensembles to achieve better performances than the one achieved by a single classifier.

Finally, in one of our previous works~\cite{giacomello2021image} we trained several CNNs on the CheXpert dataset and used them to extract image embeddings that could be later exploited as input for simpler (e.g., tree-based) classifiers.
Our results showed that image embeddings do contain relevant information about the images and allows to train tree-based classifiers with performances similar or better than the ones achieved by CNNs.

 \section{Methodology}
\label{sec:methodology}
	\subsection{CheXpert} \label{subsec:chexpert}
		CheXpert \cite{chexpert} is the largest collection of CXR images labeled and publicly available. Put together by the personnel of the Stanford Hospital between 2004 and 2017, it contains more than 200000 images from over 65000 patients. Two different PNG formats are provided, high and normal quality, with the former being 16-bit while the latter only 8-bit. A 14 class label accompanies each image, identifying what is in the image according to the radiology report. The main peculiarity of CheXpert is the automatic labeling method used, able to interpret the radiology report and extract the relevant information to assign the correct classes. \textit{Positive}, \textit{negative} and \textit{uncertain} are the possible values, according to the confidence level of the labeler, which are encoded in a vector of 14 elements, respectively, 1, 0, u. The main type of images available is frontal images, with the possibility, usually where the report was uncertain and therefore an uncertain value for the label is more likely, to find also lateral ones. Together with the vast number of automatically labeled images, the authors provide a smaller set of 200 images annotated by hand by expert radiologists, originally intended for test or validation purposes of the models.

		\begin{table}
			\begin{tabularx}{\columnwidth}{|| X | X  | X  | X  ||}
				\hline
				\textbf{Class} & \textbf{Positive(\%)} & \textbf{Uncertain(\%)} & \textbf{Negative(\%)}\\
\hline \hline
				No findings 		&16974 (8.89) 		&0 (0.0) 			&174053 (91.11)\\
				\hline
				Enlarged Card. 		&30990 (16.22)		&10017 (5.24) 		&150020 (78.53)\\
				\hline
				Cardiomegaly 		&23385 (12.24)		&549 (0.29) 		&167093 (87.47)\\
				\hline
				Lung Opacity 		&137558 (72.01	&2522 (1.32) 		&50947 (26.67) \\
				\hline
				Lung Lesion 		&7040 (3.69) 		&841 (0.44) 		&183146 (95.87)\\
				\hline
				Edema 			&49675 (26.0) 		&9450 (4.95) 		&131902 (69.05)\\
				\hline
				Consolidation 		&16870 (8.83) 		&19584 (10.25)		&154573 (80.92)\\
				\hline
				Pneumonia 		&4675 (2.45) 		&2984 (1.56) 		&183368 (95.99)\\
				\hline
				Atelectasis 		&29720 (15.56)		&25967 (13.59)		&135340 (70.85)\\
				\hline
				Pneumothorax 		&17693 (9.26)		&2708 (1.42) 		&170626 (89.32)\\
				\hline
				Pleural Effusion 	&76899 (40.26) 	&9578 (5.01) 		&104550 (54.73)\\
				\hline
				Pleural Other 		&2505 (1.31) 		&1812 (0.95) 		&186710 (97.74)\\
				\hline
				Fracture 			&7436 (3.89) 		&499 (0.26) 		&183092 (95.85)\\
				\hline
				Support Devices 	&107170 (56.1)		&915 (0.48)		&82942 (43.42)\\
				\hline
			\end{tabularx}
			\caption{CheXpert classes distribution}
			\label{table:chexpert}
		\end{table}
		 In Table \ref{table:chexpert} can be found the distribution of the data. It is important to notice that the labels are not mutually exclusive and more than one label can be present in the same image, making the classification problem a multi-label one;
		 in addition, the labels in Table \ref{table:chexpert} features a hierarchical structure.
		In order to obtain the best model possible, all these characteristic should be taken care of.
To deal with the uncertain labels included in \cite{chexpert} dataset, several policies have been proposed.
		The most simple one consists of assuming all the uncertain labels as either positive or negative.
		However, this strategy intuitively results in several wrongly labeled examples and might negatively affect the training.
		This issue was discussed in detail in \cite{pham2020} where a different strategy has been proposed, following an approach called \textit{Label-Smoothing Regularisation} -- originally introduced in \cite{szegedy2015rethinking} -- which consists of mapping the uncertain labels to random values sampled from a uniform distribution $X\sim U(\alpha,\beta)$ with $\beta > \alpha > 0.5$.
		This spares the model from using wrongly labeled examples with excessive confidence during the learning process.

	\subsection{Embeddings extraction} \label{subsec:vaeextract}
		$\beta$-VAE, introduced in \cite{Higgins2017betaVAELB} are an evolution of VAE in which $\beta$ is introduced as hyperparameter in order to balance more effectively the effect of the two components of the loss function, the reconstruction and the regularisation (actually, in this way, VAE become a particular case of $\beta$-VAE with $\beta = 1$); it is usually used to push the latent space in the direction of a more disentangled representation of the input distribution.
		$\beta$-VAE are composed of an encoder, which learns the latent representation, and a decoder, which reconstructs the input data. In this work we chose some of the most popular CNN architectures, similarly as done in \cite{giacomello2021image}, as backbones to build the encoders, trying to understand if the differences in architectures, number of parameters and characteristics significantly affect the results. The model choices are : DenseNet121, DenseNet169, DenseNet201 \cite{huang2018densely}, Xception \cite{chollet2017xception} and InceptionResNetv2 \cite{szegedy2016inceptionv4}. The encoder has been built cutting the backbone after a Global Average Pooling (GAP) layer, hence without its densely connected layers, which have been substituted by the mean and standard deviation of the learned distribution. The weights, up until the GAP layer, are the same of the models in \cite{giacomello2021image}, pre-trained on CheXpert, and are not updated during the training; only the weights after the GAP in the encoder and the decoder ones are learned. Two sizes of the latent space have been tested: 100 and 200 unit vectors. The decoder is built in the same way for each model, in order to better compare the results, and features an initial 512 units densely connected layer which is then reshaped and passed through seven convolutional blocks composed of convolution (with a number of filters halved after each block, starting from 256) and upsampling. Neither fine tuning nor any transfer learning is applied to any decoder, weights are learned from scratch for each encoder.
		Autoencoders are trained in an unsupervised way because they are able to learn the relevant feature to reconstruct the input data, therefore labels are not considered. In this sense, the main idea of using $\beta$-VAE is basically to enhance the use of CNN as universal feature extractors, so that the extracted features are both informative and general.

	\subsection{ML Models and Ensembles} \label{subsec:treembeddings}
		Once the features have been extracted, another model is used to classify them in order to demonstrate how informative they are. To serve this purpose, four ML kind of models have been tested : Random Forest (RF) \cite{tim1995random}, Gradient Boosting (GB) \cite{friedman2000greedy}, Extremely Randomized Trees (XRT) \cite{journals/ml/GeurtsEW06}, which are all tree-based, and K-Nearest Neighbours (KNN) \cite{thomas2016knn}. The choice has gone in the direction of finding something which does not requires extensive training sessions or powerful hardware to be trained, but is still able to achieve good performances. In this respect, tree based classifier ensembles have been proved to be of great value.
		Moreover, the prediction have been combined to improve the classification results, in two ways which follow what done in \cite{giacomello2021image}:
		\begin{itemize}
			\item{Simple Average:} the prediction are evenly averaged; being $y_i(\pmb{x})$ the probabilities from a certain ML model trained on the embeddings extracted from a single $\beta$-VAE $i$, the final predictions are:
				\begin{equation} \label{eq:average}
					\widetilde{y}(\pmb{x}) = \frac{1}{N}\sum_{i=1}^{N}{y_i(\pmb{x})}
				\end{equation}

			\item{Entropy-Weighted Average:} the probabilities are averaged with weights representing the confidence in such predictions.
			A possible approach to compute such confidence, is based on \textit{entropy}, which in Information Theory is defined as the uncertainty of the outcomes of random variables.
			Therefore, modeling the prediction of each classifier as a random variable following a Bernoulli distribution (the disease is either present or not), we can consider the entropy of such variable as a direct measure of the confidence of the classifier in that prediction; setting $p_{i, k}$ as the prediction for classifier $i$ (as before) and class $k$, we have that :
				\begin{equation} \label{eq:entropy}
					H_k(y_{i, k}) = -y_{i, k}log_2(y_{i, k}) - (1-y_{i, k})log_2(1-y_{i, k})
				\end{equation}
				Here, $H_k(y_{i, k}) $ is the uncertainty linked to classifier $i$ for label $k$, therefore we can see $1-H_k(y_{i, k})$ as the confidence in prediction $y_{i, k}$. The natural use of this, is as weight for averaging the single classifier prediction, so that predictions which the model is \textit{mode sure} of are weighted more while lowest scores are considered less:
				\begin{equation} \label{eq:entropyaverage}
					\widetilde{y}(\pmb{x}) = \frac{1}{N}\sum_{i=1}^{N}{(1-H_k(y_{i, k}))y_{i, k}}
				\end{equation}
		\end{itemize}
		Another option would be to consider stacking \cite{wolpert1992stacking}, but in \cite{giacomello2021image} it has been evidenced how it actually yields worse predictions compared to the simpler ensemble strategies described above.
 \section{Experimental Design}
\label{sec:design}
	\subsection{Preprocessing} \label{subsec:preprocessing}
		We trained our $\beta$-VAE only on the frontal images included in CheXpert because they are the vast majority and, usually, are the only exam performed; lateral images are there to help solve some uncertainties in the diagnosis, therefore using them would make the approach less general. Dataset has been split as follows:
		\begin{itemize}
			\item \textbf{Training} : 90\% of the dataset ($N = 189116$ examples)
			\item \textbf{Validation} : 10\% of the dataset($N = 1911$ examples)
			\item \textbf{Test} : an additional small dataset provided together with CheXpert \cite{chexpert}, composed of $N = 202$ examples of images manually annotated by experts
		\end{itemize}
		All the additional data other than the frontal images and their labels have been ignored in this work. Images have been resized to 256x256, then a template matching has been applied in order to crop them and keep the relevant part only, trying to exclude artifacts and non-random noise, such as texts usually printed together with the radiography; the final image size is then set to 224x224 and two channels have been added to make them actually 224x224x3 to make them compatible with the backbones used for the $\beta$-VAE. Contrarily to what has been done in \cite{giacomello2021image} and \cite{pham2020}, since we are not starting the training from ImageNet \cite{imagenet} weights, the images have not been normalized with ImageNet mean and standard deviation.

	\subsection{$\beta$-VAE training} \label{subsec:vaetraining}
		As described in \ref{subsec:vaeextract} , five pre-trained backbones have been used to build and train the $\beta$-VAE models, retaining the convolutional layers and discarding the densely connected classification layers, substituted by the ones yielding the mean and the logarithm of the standard deviation (it is common to use the logarithm to avoid getting too small values).

		\begin{figure}
			\centering
			\includegraphics[width=0.6\columnwidth]{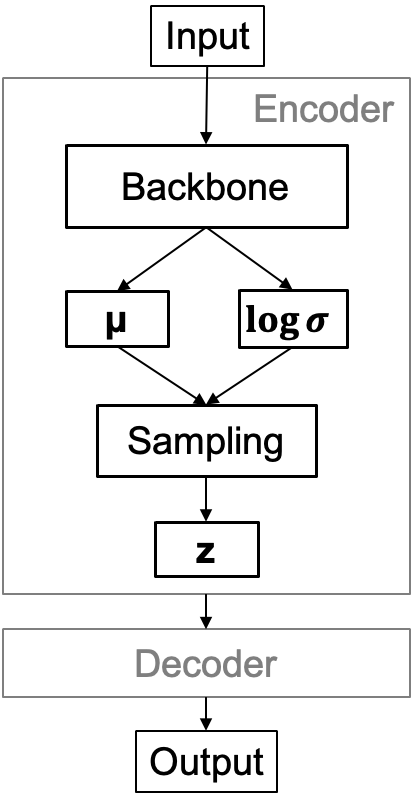}
			\caption{A block scheme of the $\beta$-VAE built for our work}
			\label{fig:vae}
		\end{figure}

		A scheme of the whole models can be found in Figure \ref{fig:vae}. Tensorflow 2.0 API \cite{tensorflow2015-whitepaper} has been used to build and train the models, with the built-in Keras modules \cite{chollet2015}. Hyperparameters and other relevant training details used for the training are described below.

			\medskip\noindent
			\textbf{Loss function.} In $\beta$-VAE the loss is composed by two components. The first component is the reconstruction loss computed as the sum of the absolute difference between the reconstructed and the original input ($\pmb{z} = g(\pmb{x})$ and $\pmb{x_{rec}} = f(\pmb{z})$), defined as:
				 \begin{equation}
					 \begin{aligned}
						L_{rec}(\phi) = -E_{z \sim q_\phi(\pmb{z}|\pmb{x})}logp_\theta(\pmb{x}|\pmb{z})
						\\= \sum_{i=1}^{N}|x_i - f_\theta(g_\phi(x_i))|.
					\end{aligned}
					\label{eq:recloss}
				\end{equation}
				The second component is the regularization, that use Kullback-Leibler divergence to force the learned distribution to fit a Normal one:
				\begin{equation}
					\begin{aligned}
						L_{KL}(\beta) =\beta D_{KL}(q_\phi(\pmb{z}|\pmb{x})||p_\theta(\pmb{z}))
						\\ =-0.5 \sum_{i=1}^{N}(1 + log(z_{\sigma i}) - z_{\mu i}^2 - e^{log(z_{\sigma i})}),
					\end{aligned}
					\label{eq:regloss}
				\end{equation}
				where $\pmb{z}_\mu$ is the mean of the latent distribution and $\pmb{z}_\sigma$ is the logarithm of its standard deviation.
				Thus, Equation \eqref{eq:recloss} and Equation \eqref{eq:regloss} are combined through $\beta$:
				\begin{equation}
					L(\phi, \theta) = L_{rec} + \beta L_{KL}
				\label{eq:loss}
				\end{equation}

			\medskip\noindent
			\textbf{Coefficient $\pmb{\beta}$.}
			After several trials with different values, we found that it is better to use a scheduler that progressively increases its value, in order to allow the models to learn to reconstruct the data without being overwhelmed by the otherwise dominant latent loss, which forces the model to fit a normal distribution which does not have any value for reconstruction. The schedule starts with three epochs with $\beta = 0$, increasing its value progressively as $0.005 * 1.2^epoch$.

			\medskip\noindent
			\textbf{Optimizer.}
			Adam algorithm \cite{kingma2017adam} has been used to optimize the weights after each forward pass, with the default parameters from the Tensorflow built in module.

			\medskip\noindent
			\textbf{Learning rate.}
			We used a decreasing learning rate schedule when the validation loss function, computed over the validation set described in \ref{subsec:preprocessing} at the end of each epoch, stopped decreasing, starting from $7.5*10^{-4}$ and halving whenever needed.

			\medskip\noindent
			\textbf{Epochs.}
			We fine tuned our models for 10 epochs each.

	\subsection{ML Classifiers training} \label{subsec:treetraining}
		After training the $\beta$-VAE they have been used to extract embeddings on which ML models would be trained. To do so, we passed the validation set to each of VAE to create datasets made of embeddings, to be fed to the ML models. In order to tune the many hyperparameters of these models, we used a grid search approach on several of them. In particular, for tree based architectures we optimised the maximum depth of the trees, which balances accuracy and overfitting tendencies; the minimum number of samples to allow the split of an internal tree node; the minimum number of samples required for a node to become a leaf; the number of estimators in the ensembles.

		\begin{table}
			\begin{tabularx}{\columnwidth}{|| X || Y | Y | Y | Y ||}
				\hline
				\textbf{Model} & \textbf{Estimators} & \textbf{Max depth} & \textbf{Min sample leaf} & \textbf{Min sample split}\\
				\hline \hline
				RF			&2000	&10		&2			&2\\
				\hline
				XRT	&2000	&10		&1			&5\\
				\hline
				GB			&1000	&3		&default		&default\\
				\hline
			\end{tabularx}
			\caption{Parameters used for tree-based classifiers after grid search. \emph{Estimators} is the number of estimators in the ensemble; \emph{Max depth} indicates the maximum depth of the trees; \textit{Min sample leaf} is the minimum samples needed to allow a split; \emph{Min sample leaf} represents the minimum samples required for a node to become a leaf; \emph{Default}  means that the default SciKit Learn value has been used}
			\label{table:params}
		\end{table}

		Table~\ref{table:params} show the results of parameters tuning for tree-based methods; for KNN we
		  tuned the number of neighbours, achieving best performances with $k=10$ .
		To train the ML models, the dataset built from embeddings extracted from the validation set have been used; the same extraction process has been applied to the test dataset.To build and train the ML models, \textit{SciKit Learn} \cite{scikit-learn} library has been used.

	\subsection{Performance Evaluation}\label{subsec:perfeval}
		Performance of the $\beta$-VAE can be difficult to analyze because it is not always clear how to state what a good reconstruction is, so a qualitative analysis of the generated images compared to original ones is often the way to go.
		Concerning the classification problems, the metric used is the Area Under the Receiving Operating Characteristic (AUROC), computed plotting the True Positive Rate (TPR) against the False Positive Rate (FPR); for multi-class and multi-label problems it has to be done in a one-vs-all fashion. In \cite{mandrekar2010auroc} it is evidenced how values $>0.7$ are considered acceptable for the medical field, $>0.8$ point to very good results and $>0.9$ an outstanding indicator.
 \section{Results}
\label{sec:results}
	In the first part of this section the focus will be on $\beta$-VAE, to show some examples of their reconstruction capabilities; in the following parts, otherwise, the main topic will shift on the informative value of the embeddings, showing the capabilities of the rather simple ML models trained on them; the AUROC scores have been tested on five classes only (cardiomegaly, edema, atelectasis, pleural effusion, consolidation) because they are the most represented in the test dataset while they are also the same used in other work such as \cite{chexpert}, \cite{pham2019interpreting} and \cite{giacomello2021image}.
	\subsection{VAE generated images} \label{subsec:vaegen}
		As mentioned in \ref{subsec:perfeval}, the main aspect related to the quality of an autoencoder model, in particular when dealing with convolutional models, it is how realistic the reconstructed images look. In general, images reconstructed from a feature vector in an autoencoder tend to be blurry and therefore not so realistic from a human observer standpoint, but the main characteristics of the original images (shape, color, details) are clearly visible, meaning that the encoder was able to capture the most relevant aspects in the feature vector.
\begin{figure*}
   			 \centering
   			 \includegraphics[width=0.8\textwidth]{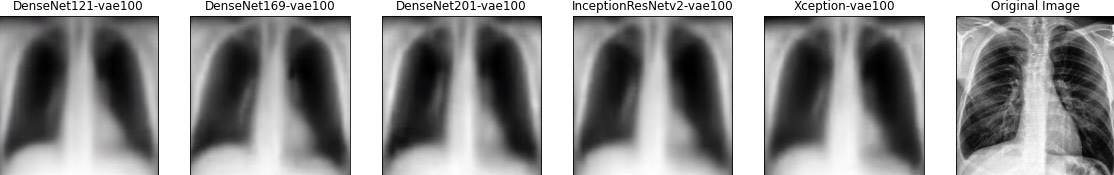}
			 \\[\smallskipamount]
			 \includegraphics[width=0.8\textwidth]{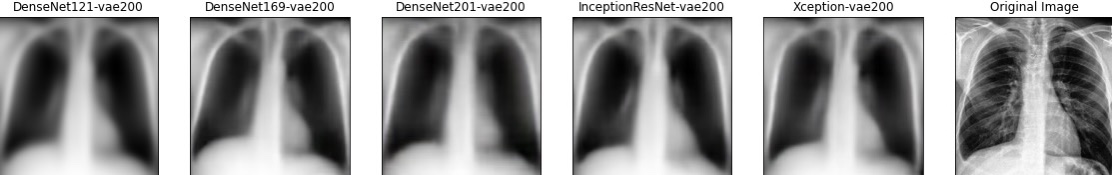}
			 \\[\smallskipamount]
   			 \includegraphics[width=0.8\textwidth]{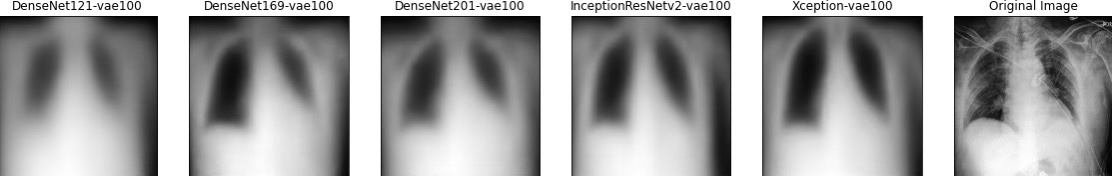}
			 \\[\smallskipamount]
			 \includegraphics[width=0.8\textwidth]{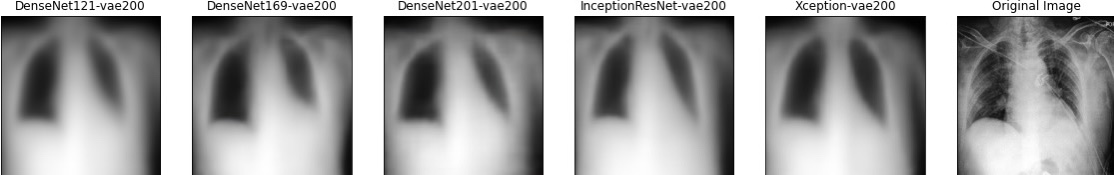}
			 \caption{Output images from $\beta$-VAE with 100 ($1^{st}$ and $3^{rd}$ row) and 200 ($2^{nd}$ and $4^{th}$ row) units latent space, given as input the same two images. The label identifies the backbone used for the encoder. It can be noted that the images have subtle differences between them but they all clearly are a blurred and less detailed version of the original image}
   			 \label{fig:vae}
		\end{figure*}
This is definitely the case for our models, as it can be seen in fig. \ref{fig:vae} where two examples from CheXpert and their reconstruction through different $\beta$-VAE are shown. It is evident how the output looks exactly like a blurred, lower definition version of the input, which still is perfectly recognizable. This makes us think that, at least for what concern the encoding and decoding capabilities of our models, they are  capable of identifying the main features. From this first comparison it has to be noted also that passing from a 100 feature vector to a 200 one does not seem to improve much the reconstruction. Unexpectedly, DenseNet121 models, despite yielding, as further highlighted in the next subsection, the far worst features for predictions, do not look significantly worse for at least three reconstructions, with only a subtle more blurred effect, which however becomes very evident in the second example of models with a 100 feature vector, where the \textit{silhouette} of the original image is almost completely lost.  This more accentuated blurring, even if very subtle in most of the images, possibly means a less precise encoded information, resulting later in poorer classification capabilities. 

	\subsection{Predictions from single VAE} \label{subsec:vaepred}
		Once the VAE have been trained, they have been used to generate the embeddings on which four kind of ML classifiers have been trained.

		\begin{table*}
			\centering
			\begin{tabularx}{\textwidth}{|| X || Y | Y | Y | Y | Y || Y ||}
				\hline
				\textbf{Model-100 units}         & \textbf{Cardiomegaly} & \textbf{Edema} & \textbf{Consolidation} & \textbf{Atelectasis} & \textbf{Pleural Effusion} & \textbf{Mean}  \\
				\hline \hline
				DNet121-GB    & 0.555                & 0.669                & 0.533                & 0.528                & 0.612                & 0.580                \\
				DNet121-XRT   & {\ul 0.618}          & 0.706                & {\ul 0.741}          & {\ul 0.667}          & {\ul 0.753}          & {\ul 0.697}          \\
				DNet121-KNN   & 0.511                & 0.681                & 0.552                & 0.545                & 0.649                & 0.587                \\
				DNet121-RF    & 0.591                & {\ul 0.726}          & 0.702                & 0.650                & 0.742                & 0.682                \\
				\hline \hline
				DNet169-GB    & 0.634                & 0.694                & 0.424                & 0.607                & 0.752                & 0.622                \\
				DNet169-XRT   & {\ul 0.798}          & {\ul 0.815}          & {\ul 0.845}          & 0.735                & {\ul 0.866}          & {\ul 0.812}          \\
				DNet169-KNN   & 0.587                & 0.724                & 0.638                & 0.598                & 0.815                & 0.673                \\
				DNet169-RF    & 0.797                & {\ul 0.815}          & 0.830                & {\ul \textbf{0.751}} & 0.864                & 0.811                \\
				\hline \hline
				DNet201-GB    & 0.605                & 0.619                & 0.637                & 0.559                & 0.732                & 0.630                \\
				DNet201-XRT   & {\ul 0.749}          & {\ul 0.798}          & {\ul \textbf{0.863}} & 0.720                & 0.854                & {\ul 0.797}          \\
				DNet201-KNN   & 0.586                & 0.717                & 0.585                & 0.650                & 0.795                & 0.667                \\
				DNet201-RF    & 0.732                & 0.785                & 0.849                & {\ul 0.722}          & {\ul 0.855}          & 0.789                \\
				\hline \hline
				IncResNet-GB  & 0.615                & 0.733                & 0.646                & 0.604                & 0.789                & 0.677                \\
				IncResNet-XRT & {\ul \textbf{0.805}} & 0.796                & {\ul 0.862}          & {\ul 0.746}          & 0.872                & {\ul \textbf{0.816}} \\
				IncResNet-KNN & 0.669                & 0.752                & 0.621                & 0.643                & 0.809                & 0.699                \\
				IncResNet-RF  & 0.787                & {\ul 0.801}          & 0.826                & 0.740                & {\ul \textbf{0.873}} & 0.806                \\
				\hline \hline
				Xception-GB   & 0.656                & 0.612                & 0.554                & 0.615                & 0.705                & 0.628                \\
				Xception-XRT  & {\ul 0.760}          & {\ul \textbf{0.820}} & {\ul 0.841}          & {\ul 0.724}          & {\ul 0.819}          & {\ul 0.793}          \\
				Xception-KNN  & 0.637                & 0.737                & 0.640                & 0.527                & 0.752                & 0.659                \\
				Xception-RF   & 0.749                & 0.816                & 0.827                & 0.721                & 0.818                & 0.786\\
\hline \hline \hline
				\textbf{Model-200 units}         & \textbf{Cardiomegaly} & \textbf{Edema} & \textbf{Consolidation} & \textbf{Atelectasis} & \textbf{Pleural Effusion} & \textbf{Mean}  \\
				\hline \hline
				DNet121-GB	& 0.534					& 0.592				& 0.574				& 0.521					& 0.701					& 0.585 \\
				DNet121-XRT	& 0.717					& \underline{0.708} 		& 0.614				& \underline{0.708}			& 0.754					& \underline{0.700} \\
				DNet121-KNN	& 0.581					& 0.511				& 0.517				& 0.554					& 0.598					& 0.552 \\
				DNet121-RF	& \underline{0.726}			& 0.704				& \underline{0.618}		& 0.672					& \underline{0.760}			& 0.696 \\
				\hline \hline
				DNet169-GB	& 0.534					& 0.737				& 0.493				& 0.498					& 0.772					& 0.607 \\
				DNet169-XRT	& \underline{0.799}			& \underline{0.810}		& \underline{0.881}		& 0.772					& 0.877					& \textbf{\underline{0.828}} \\
				DNet169-KNN	& 0.617					& 0.729				& 0.624				& 0.637					& 0.841					& 0.690 \\
				DNet169-RF	& 0.786					& 0.801				& 0.860				& \textbf{\underline{0.791}}	& \textbf{\underline{0.878}}	& 0.823 \\
				\hline \hline
				DNet201-GB	& 0.439					& 0.713 				& 0.432				& 0.584					& 0.778					& 0.589 \\
				DNet201-XRT	& \underline{0.751}			& 0.807 				& 0.880				& \underline{0.726}			& 0.859					& \underline{0.804} \\
				DNet201-KNN	& 0.625					& 0.714 				& 0.668				& 0.557					& 0.821					& 0.677 \\
				DNet201-RF	& 0.725					& \underline{0.808} 		& \textbf{\underline{0.899}} & 0.709					& \underline{0.864}			& 0.801 \\
				\hline \hline
				IncResNet-GB	& 0.529					& 0.654 				& 0.423				& 0.517					& 0.754					& 0.575 \\
				IncResNet-XRT	& 0.810					& 0.800 				& \underline{0.897}		& 0.741					& 0.867					& 0.823 \\
				IncResNet-KNN& 0.708					& 0.771 				& 0.638				& 0.551					& 0.863					& 0.706 \\
				IncResNet-RF	& \textbf{\underline{0.816}}	& \underline{0.803} 		& 0.891				& \underline{0.744}			& \underline{0.868}			& \underline{0.824} \\
				\hline \hline
				Xception-GB	& 0.470					& 0.643				& 0.592				& 0.513					& 0.792					& 0.602 \\
				Xception-XRT	& \underline{0.800}			& \textbf{\underline{0.816}}& \underline{0.860}		& \underline{0.748}			& \underline{0.852}			& \underline{0.815} \\
				Xception-KNN	& 0.707					& 0.760				& 0.581				& 0.627					& 0.791					& 0.693 \\
				Xception-RF	& 0.799					& 0.814				& 0.843				& 0.728					& 0.851					& 0.807\\
				\hline
			\end{tabularx}
			\caption{AUROC scores obtained from ML models trained on embeddings extracted form single $\beta$-VAE, considered label-wise and as an average (Mean column), rounded to the third decimal; the first and second half shows results obtained on models with respectively 100 and 200 units latent vector; the best result for the single $\beta$-VAE for a specific class is {\ul underlined}, the best overall for models with the same feature vector size is in \textbf{bold}.}
			\label{table:singleresult}	
		\end{table*}

		In table \ref{table:singleresult} the AUROC scores obtained are shown. It is evident how there are some very relevant differences on the performances, with KNN being completely unable to yield reliable predictions. XRT here looks like the one offering the best performances, closely followed by RF, while GB, despite showing an acceptable result according to \cite{mandrekar2010auroc} , performs significantly worse. Considering the results of RF and XRT, embedding  are proved to be informative enough to be used for robust classification tasks, with their best performances steadily above 80\% AUROC except for DenseNet201 and Xception, but only for 100 feature vector. It has to be noted how DenseNet121 in both cases seems to give significantly less informative features, causing all the models to achieve consistently and evidently worse results than both its siblings 169 and 201, but also Inception and Xception; this possibly means that the 121 architecture is not suited to extract feature from this kind of images. Label wise, they all seem to follow the same pattern, with \textit{atelectasis} being almost always the class with the worse score, while the models have much better confidence in spotting \textit{consolidation} and \textit{pleural effusion}, often reaching close to 90\%. Differences among other backbones with the same ML models are not very important as they remain under 2-3\%. If we then consider the different size of the feature vector, we can observe how, quite predictably, ML models trained on bigger feature vectors score consistently better than the other, even though the difference is not large, again around 2\% both for single labels and overall.

		\begin{table*}
			\centering
			\begin{tabularx}{\textwidth}{|| X || Y | Y | Y | Y | Y || Y ||}
				\hline
				\textbf{Model 100}         & \textbf{Cardiomegaly} & \textbf{Edema} & \textbf{Consolidation} & \textbf{Atelectasis} & \textbf{Pleural Effusion} & \textbf{Mean}  \\
				\hline \hline
				Avg-GB			& 0.722				& 0.751			& 0.539			& 0.640			& 0.802			& 0.691 \\
				Entropy Avg-GB	& 0.722				& 0.752			& 0.539			& 0.640			& 0.800			& 0.690 \\
				\hline \hline
				Avg-RF			& 0.782				& \textbf{0.817}		& 0.858			& \textbf{0.757}		& \textbf{0.855}		& 0.814 \\
				Entropy Avg-RF	& 0.776				& 0.801			& 0.858			& 0.752			& 0.852			& 0.808 \\
				\hline \hline
				Avg-XRT			& \textbf{0.797}			& 0.815			& \textbf{0.865}		& 0.744			& \textbf{0.855}		& \textbf{0.815} \\
				Entropy Avg-XRT	& 0.795				& 0.800			& \textbf{0.865}		& 0.744			& 0.852			& 0.811\\
				\hline \hline \hline
				\textbf{Model 200}         & \textbf{Cardiomegaly} & \textbf{Edema} & \textbf{Consolidation} & \textbf{Atelectasis} & \textbf{Pleural Effusion} & \textbf{Mean}  \\
				\hline \hline
				Avg-GB			& 0.623				& 0.786			& 0.526			& 0.544			& 0.840			& 0.664 \\
				Entropy Avg-GB	& 0.623				& 0.785			& 0.526			& 0.544			& 0.839			& 0.663 \\
				\hline \hline
				Avg-RF			& 0.814				& 0.813			& 0.883			& 0.773			& 0.878			& 0.832 \\
				Entropy Avg-RF	& 0.815				& 0.802			& 0.884			& 0.776			& \textbf{0.880}		& 0.831 \\
				\hline \hline
				Avg-XRT			& \textbf{0.819}			& \textbf{0.814}		& \textbf{0.895}		& 0.775			& 0.871			& \textbf{0.835} \\
				Entropy Avg-XRT	& 0.818				& 0.804			& 0.894			& \textbf{0.777}		& 0.872			& 0.833\\
				\hline
			\end{tabularx}
			\caption{ AUROC scores obtained from ML models ensembling, split for feature vector (100 above, 200 below). Again, single class and overall results are presented. The best value in the column is highlighted in bold for both the halves.}
			\label{table:ensembleresult}
		\end{table*}
	
	\subsection{Ensemble Predictions} \label{subsec:enspred}
		To improve the AUROC score virtually \textit{for free}, meaning that no further training, feature engineering, or parameters tuning would be required in a real world scenario, two ensembling techniques have been tested to combine the results, as explained in \ref{subsec:treembeddings}. Again, in table \ref{table:ensembleresult} the results obtained are shown. The main advantage about using entropy averaged prediction is the fact that if models have different strengths on different classes, these can be captured by the entropy based weights and yield a better overall result, as it seems to be happening in \cite{giacomello2021image} which uses the values extracted from GAP layers of single CNN as embeddings. The results of these ensembles seems to confirm this observation, since both averaging methods lead to a consistent improvement on the single scores, but there is little to no difference between entropy weighted and simple average; as noted in the previous section, it is evident how the ML models trained on single VAE embeddings seem to have the same strengths and weaknesses and this cancels any advantage of using such a weighting strategy. Perhaps, more sophisticated ensemble strategies might help to improve this result, but this is beyond the scope of this work.
		Once again, exactly like on single models, ensembles on bigger feature vectors have a better AUROC, matching the improvements.
 \section{Conclusion}
\label{sec:conclusions}
	In this work we trained $\beta$-Variational Autoencoders on the frontal CXR images from the CheXpert public dataset, using several different CNN architectures as base building block. Our interest was to automatically extract the most general possible high level features containing sufficient information to be used to classify them according to the disease present in the original image. We used Machine learning tree-based and neighbours-based models to classify such images, in order to assess the usability of the feature extracted for such a task. The results obtained are very promising because, despite being inferior to state of the art classifiers trained specifically for classification, because they still output very reliable predictions, according to their AUROC scores. Model ensembling, based on averaging the prediction with or without weights, have been used to improve the results without the necessity to undergo further training, model tuning or feature engineering, with consistent success over both single classes and overall results. 
	In conclusion, our results suggest that the features extracted by the autoencoders are indeed highly informative and allow to classify the images. Nonetheless, further investigations will be necessary to test if these features are general enough to be used also on slightly different tasks with the respect to the one considered in this work, in particular evaluating them on images or dataset from different sources and with different classes. Also, more specific experiments on the feature vector size should be conducted to find the optimal dimension able to capture the necessary information to achieve even better classification performances while keeping the encoding capabilities general. Finally, despite simpler ensembling methods based on averaging the predictions proved successful in our experimental analysis, more sophisticated approaches, such as stacked generalization, might be evaluated as well in future works.

\bibliographystyle{IEEEtran}

\begin{thebibliography}{}
\providecommand{\url}[1]{#1}
\csname url@samestyle\endcsname
\providecommand{\newblock}{\relax}
\providecommand{\bibinfo}[2]{#2}
\providecommand{\BIBentrySTDinterwordspacing}{\spaceskip=0pt\relax}
\providecommand{\BIBentryALTinterwordstretchfactor}{4}
\providecommand{\BIBentryALTinterwordspacing}{\spaceskip=\fontdimen2\font plus
\BIBentryALTinterwordstretchfactor\fontdimen3\font minus
  \fontdimen4\font\relax}
\providecommand{\BIBforeignlanguage}[2]{{%
\expandafter\ifx\csname l@#1\endcsname\relax
\typeout{** WARNING: IEEEtran.bst: No hyphenation pattern has been}%
\typeout{** loaded for the language `#1'. Using the pattern for}%
\typeout{** the default language instead.}%
\else
\language=\csname l@#1\endcsname
\fi
#2}}
\providecommand{\BIBdecl}{\relax}
\BIBdecl

\end{thebibliography}


\begin{thebibliography}{10}
\providecommand{\url}[1]{#1}
\csname url@samestyle\endcsname
\providecommand{\newblock}{\relax}
\providecommand{\bibinfo}[2]{#2}
\providecommand{\BIBentrySTDinterwordspacing}{\spaceskip=0pt\relax}
\providecommand{\BIBentryALTinterwordstretchfactor}{4}
\providecommand{\BIBentryALTinterwordspacing}{\spaceskip=\fontdimen2\font plus
\BIBentryALTinterwordstretchfactor\fontdimen3\font minus
  \fontdimen4\font\relax}
\providecommand{\BIBforeignlanguage}[2]{{\expandafter\ifx\csname l@#1\endcsname\relax
\typeout{** WARNING: IEEEtran.bst: No hyphenation pattern has been}\typeout{** loaded for the language `#1'. Using the pattern for}\typeout{** the default language instead.}\else
\language=\csname l@#1\endcsname
\fi
#2}}
\providecommand{\BIBdecl}{\relax}
\BIBdecl

\bibitem{lecun}
Y.~{LeCun}, B.~{Boser}, J.~S. {Denker}, D.~{Henderson}, R.~E. {Howard},
  W.~{Hubbard}, and L.~D. {Jackel}, ``Backpropagation applied to handwritten
  zip code recognition,'' \emph{Neural Computation}, vol.~1, no.~4, pp.
  541--551, Dec 1989.

\bibitem{vgg16}
K.~Simonyan and A.~Zisserman, ``Very deep convolutional networks for
  large-scale image recognition,'' 2014.

\bibitem{greenspan2016guest}
H.~Greenspan, B.~van Ginneken, and R.~M. Summers, ``Guest editorial deep
  learning in medical imaging: Overview and future promise of an exciting new
  technique,'' \emph{IEEE Transactions on Medical Imaging}, vol.~35, no.~5, pp.
  1153--1159, 2016.

\bibitem{ravi2017deep}
D.~Ravì, C.~Wong, F.~Deligianni, M.~Berthelot, J.~Andreu-Perez, B.~Lo, and
  G.-Z. Yang, ``Deep learning for health informatics,'' \emph{IEEE Journal of
  Biomedical and Health Informatics}, vol.~21, no.~1, pp. 4--21, 2017.

\bibitem{giacomello2021image}
E.~Giacomello, P.~L. Lanzi, D.~Loiacono, and L.~Nassano, ``Image embedding and
  model ensembling for automated chest x-ray interpretation,'' 2021.

\bibitem{irvin2019chexpert}
J.~Irvin, P.~Rajpurkar, M.~Ko, Y.~Yu, S.~Ciurea-Ilcus, C.~Chute, H.~Marklund,
  B.~Haghgoo, R.~Ball, K.~Shpanskaya, J.~Seekins, D.~A. Mong, S.~S. Halabi,
  J.~K. Sandberg, R.~Jones, D.~B. Larson, C.~P. Langlotz, B.~N. Patel, M.~P.
  Lungren, and A.~Y. Ng, ``Chexpert: A large chest radiograph dataset with
  uncertainty labels and expert comparison,'' 2019.

\bibitem{chexpert_competition}
S.~M. Group, ``Chexpert competition,''
  \url{https://stanfordmlgroup.github.io/competitions/chexpert/}.

\bibitem{pham2019interpreting}
H.~H. Pham, T.~T. Le, D.~Q. Tran, D.~T. Ngo, and H.~Q. Nguyen, ``Interpreting
  chest x-rays via cnns that exploit disease dependencies and uncertainty
  labels,'' 2019.

\bibitem{johnson2019mimiccxrjpg}
A.~E.~W. Johnson, T.~J. Pollard, N.~R. Greenbaum, M.~P. Lungren, C.~ying Deng,
  Y.~Peng, Z.~Lu, R.~G. Mark, S.~J. Berkowitz, and S.~Horng, ``Mimic-cxr-jpg, a
  large publicly available database of labeled chest radiographs,'' 2019.

\bibitem{Wang_2017}
\BIBentryALTinterwordspacing
X.~Wang, Y.~Peng, L.~Lu, Z.~Lu, M.~Bagheri, and R.~M. Summers, ``Chestx-ray8:
  Hospital-scale chest x-ray database and benchmarks on weakly-supervised
  classification and localization of common thorax diseases,'' \emph{2017 IEEE
  Conference on Computer Vision and Pattern Recognition (CVPR)}, Jul 2017.
  [Online]. Available: \url{http://dx.doi.org/10.1109/CVPR.2017.369}
\BIBentrySTDinterwordspacing

\bibitem{rajpurkar2017chexnet}
P.~Rajpurkar, J.~Irvin, K.~Zhu, B.~Yang, H.~Mehta, T.~Duan, D.~Ding, A.~Bagul,
  C.~Langlotz, K.~Shpanskaya, M.~P. Lungren, and A.~Y. Ng, ``Chexnet:
  Radiologist-level pneumonia detection on chest x-rays with deep learning,''
  2017.

\bibitem{densenet}
G.~Huang, Z.~Liu, L.~van~der Maaten, and K.~Q. Weinberger, ``Densely connected
  convolutional networks,'' 2016.

\bibitem{rajpurkar2018deep}
P.~Rajpurkar, J.~Irvin, R.~L. Ball, K.~Zhu, B.~Yang, H.~Mehta, T.~Duan,
  D.~Ding, A.~Bagul, C.~P. Langlotz \emph{et~al.}, ``Deep learning for chest
  radiograph diagnosis: A retrospective comparison of the chexnext algorithm to
  practicing radiologists,'' \emph{PLoS medicine}, vol.~15, no.~11, p.
  e1002686, 2018.

\bibitem{kumar2018boosted}
P.~Kumar, M.~Grewal, and M.~M. Srivastava, ``Boosted cascaded convnets for
  multilabel classification of thoracic diseases in chest radiographs,'' in
  \emph{Image Analysis and Recognition}, A.~Campilho, F.~Karray, and B.~ter
  Haar~Romeny, Eds.\hskip 1em plus 0.5em minus 0.4em\relax Cham: Springer
  International Publishing, 2018, pp. 546--552.

\bibitem{Lu2020Multi}
Z.~{Lu}, I.~{Whalen}, Y.~{Dhebar}, K.~{Deb}, E.~{Goodman}, W.~{Banzhaf}, and
  V.~N. {Boddeti}, ``Multi-objective evolutionary design of deep convolutional
  neural networks for image classification,'' \emph{IEEE Transactions on
  Evolutionary Computation}, pp. 1--1, 2020.

\bibitem{ye2020weakly}
W.~Ye, J.~Yao, H.~Xue, and Y.~Li, ``Weakly supervised lesion localization with
  probabilistic-cam pooling,'' 2020.

\bibitem{cam_paper}
\BIBentryALTinterwordspacing
R.~R. Selvaraju, M.~Cogswell, A.~Das, R.~Vedantam, D.~Parikh, and D.~Batra,
  ``Grad-cam: Visual explanations from deep networks via gradient-based
  localization,'' \emph{International Journal of Computer Vision}, vol. 128,
  no.~2, p. 336–359, Oct 2019. [Online]. Available:
  \url{http://dx.doi.org/10.1007/s11263-019-01228-7}
\BIBentrySTDinterwordspacing

\bibitem{rubin2018large}
J.~Rubin, D.~Sanghavi, C.~Zhao, K.~Lee, A.~Qadir, and M.~Xu-Wilson, ``Large
  scale automated reading of frontal and lateral chest x-rays using dual
  convolutional neural networks,'' 2018.

\bibitem{chexpert}
\BIBentryALTinterwordspacing
J.~Irvin, P.~Rajpurkar, M.~Ko, Y.~Yu, S.~Ciurea{-}Ilcus, C.~Chute, H.~Marklund,
  B.~Haghgoo, R.~L. Ball, K.~Shpanskaya, J.~Seekins, D.~A. Mong, S.~S. Halabi,
  J.~K. Sandberg, R.~Jones, D.~B. Larson, C.~P. Langlotz, B.~N. Patel, M.~P.
  Lungren, and A.~Y. Ng, ``Chexpert: {A} large chest radiograph dataset with
  uncertainty labels and expert comparison,'' \emph{CoRR}, vol. abs/1901.07031,
  2019. [Online]. Available: \url{http://arxiv.org/abs/1901.07031}
\BIBentrySTDinterwordspacing

\bibitem{pham2020}
H.~H. Pham, T.~T. Le, D.~Q. Tran, D.~T. Ngo, and H.~Q. Nguyen, ``Interpreting
  chest x-rays via cnns that exploit hierarchical disease dependencies and
  uncertainty labels,'' 2020.

\bibitem{szegedy2015rethinking}
C.~Szegedy, V.~Vanhoucke, S.~Ioffe, J.~Shlens, and Z.~Wojna, ``Rethinking the
  inception architecture for computer vision,'' 2015.

\bibitem{Higgins2017betaVAELB}
I.~Higgins, L.~Matthey, A.~Pal, C.~P. Burgess, X.~Glorot, M.~Botvinick,
  S.~Mohamed, and A.~Lerchner, ``beta-vae: Learning basic visual concepts with
  a constrained variational framework,'' in \emph{ICLR}, 2017.

\bibitem{huang2018densely}
G.~Huang, Z.~Liu, L.~van~der Maaten, and K.~Q. Weinberger, ``Densely connected
  convolutional networks,'' 2018.

\bibitem{chollet2017xception}
F.~Chollet, ``Xception: Deep learning with depthwise separable convolutions,''
  2017.

\bibitem{szegedy2016inceptionv4}
C.~Szegedy, S.~Ioffe, V.~Vanhoucke, and A.~Alemi, ``Inception-v4,
  inception-resnet and the impact of residual connections on learning,'' 2016.

\bibitem{tim1995random}
T.~K. Ho, ``Random decision forests,'' in \emph{Proceedings of 3rd
  International Conference on Document Analysis and Recognition}, vol.~1, 1995,
  pp. 278--282 vol.1.

\bibitem{friedman2000greedy}
J.~H. Friedman, ``Greedy function approximation: A gradient boosting machine,''
  \emph{Annals of Statistics}, vol.~29, pp. 1189--1232, 2000.

\bibitem{journals/ml/GeurtsEW06}
\BIBentryALTinterwordspacing
P.~Geurts, D.~Ernst, and L.~Wehenkel, ``Extremely randomized trees.''
  \emph{Mach. Learn.}, vol.~63, no.~1, pp. 3--42, 2006. [Online]. Available:
  \url{http://dblp.uni-trier.de/db/journals/ml/ml63.html}
\BIBentrySTDinterwordspacing

\bibitem{thomas2016knn}
O.~Anava and K.~Y. Levy, ``K-nearest neighbors: From global to local,'' in
  \emph{Proceedings of the 30th International Conference on Neural Information
  Processing Systems}, ser. NIPS'16.\hskip 1em plus 0.5em minus 0.4em\relax Red
  Hook, NY, USA: Curran Associates Inc., 2016, pp. 4923--4931.

\bibitem{wolpert1992stacking}
D.~Wolpert, ``Stacked generalization,'' \emph{Neural Networks}, vol.~5, pp.
  241--259, 12 1992.

\bibitem{imagenet}
J.~Deng, W.~Dong, R.~Socher, L.-J. Li, K.~Li, and L.~Fei-Fei, ``Imagenet: A
  large-scale hierarchical image database,'' in \emph{2009 IEEE Conference on
  Computer Vision and Pattern Recognition}, 2009, pp. 248--255.

\bibitem{tensorflow2015-whitepaper}
\BIBentryALTinterwordspacing
M.~Abadi, A.~Agarwal, P.~Barham, E.~Brevdo, Z.~Chen, C.~Citro, G.~S. Corrado,
  A.~Davis, J.~Dean, M.~Devin, S.~Ghemawat, I.~Goodfellow, A.~Harp, G.~Irving,
  M.~Isard, Y.~Jia, R.~Jozefowicz, L.~Kaiser, M.~Kudlur, J.~Levenberg,
  D.~Man\'{e}, R.~Monga, S.~Moore, D.~Murray, C.~Olah, M.~Schuster, J.~Shlens,
  B.~Steiner, I.~Sutskever, K.~Talwar, P.~Tucker, V.~Vanhoucke, V.~Vasudevan,
  F.~Vi\'{e}gas, O.~Vinyals, P.~Warden, M.~Wattenberg, M.~Wicke, Y.~Yu, and
  X.~Zheng, ``{TensorFlow}: Large-scale machine learning on heterogeneous
  systems,'' 2015, software available from tensorflow.org. [Online]. Available:
  \url{https://www.tensorflow.org/}
\BIBentrySTDinterwordspacing

\bibitem{chollet2015}
F.~Chollet, ``keras,'' \url{https://github.com/fchollet/keras}, 2015.

\bibitem{kingma2017adam}
D.~P. Kingma and J.~Ba, ``Adam: A method for stochastic optimization,'' 2017.

\bibitem{scikit-learn}
F.~Pedregosa, G.~Varoquaux, A.~Gramfort, V.~Michel, B.~Thirion, O.~Grisel,
  M.~Blondel, P.~Prettenhofer, R.~Weiss, V.~Dubourg, J.~Vanderplas, A.~Passos,
  D.~Cournapeau, M.~Brucher, M.~Perrot, and E.~Duchesnay, ``Scikit-learn:
  Machine learning in {P}ython,'' \emph{Journal of Machine Learning Research},
  vol.~12, pp. 2825--2830, 2011.

\bibitem{mandrekar2010auroc}
\BIBentryALTinterwordspacing
J.~N. Mandrekar, ``Receiver operating characteristic curve in diagnostic test
  assessment,'' \emph{Journal of Thoracic Oncology}, vol.~5, no.~9, pp.
  1315--1316, 2010. [Online]. Available:
  \url{https://www.sciencedirect.com/science/article/pii/S1556086415306043}
\BIBentrySTDinterwordspacing

\end{thebibliography}
\iffalse
\vspace{12pt}
\color{red}
IEEE conference templates contain guidance text for composing and formatting conference papers. Please ensure that all template text is removed from your conference paper prior to submission to the conference. Failure to remove the template text from your paper may result in your paper not being published.
\fi

\end{document}